# Raman scattering study of InAs nanowire under high pressure

Dipanwita Majumdar,[1] Abhisek Basu,[2] Goutam Dev Mukherjee,[2] Daniele Ercolani,[3] Lucia Sorba,[3] Achintya Singha,[1]*

[1]Department of Physics, Bose Institute, 93/1, Acharya Prafulla Chandra Road, Kolkata 700 009, India

[2]Department of Physics, IISER Kolkata

[3]NEST-Istituto Nanoscienze-CNR and Scuola Normale Superiore, Piazza S. Silvestro 12, I-56127 Pisa, Italy

*To whom correspondence should be addressed: E-mail: achintya@jcbose.ac.in (A. Singha), Phone: + 91 33 23031177, Fax: + 91 33 23506790

**ABSTRACT**

The pressure dependent phonon modes of predominant wurtzite InAs nanowires has been investigated in a diamond anvil cell under hydrostatic pressure up to ~ 58 GPa. The TO and LO at Γ point and other optical phonon frequencies increase linearly while the LO-TO splitting decreases with pressure. The recorded Raman modes have been used to determine the mode Grüneisen parameters and also the value of Born's transverse effective charge ($e^*_T$). The calculated $e^*_T$ exhibits a linear reduction with increasing pressure implying an increase in covalency of nanowires under compression. The intensity of the Raman modes shows a strong enhancement as the energy of $E_1$ band gap approaches the excitation energy, which has been discussed in terms of resonant Raman scattering. An indication of structural phase transformation has been observed above pressure 10.87 GPa. We propose this transformation may be from wurtzite to rock salt phase although further experimental and theoretical confirmations are needed.

**INTRODUCTION**

In recent years, one dimensional semiconductor nanowires (NWs) of diameter in the range of few to several tens of nanometer, find their potential as one of the next generation building block for nanoscale electronics,[1,2] photonics,[3] sensors,[4] and lasers.[5] Among them, InAs NWs are of particular interest for their excellent electron transport property due to remarkably high bulk mobility resulting from small effective mass. Thermodynamic parameters play an important role in structural phase change, which has a large impact on its physical properties. By applying pressure, the electronic and structural properties of semiconductors can be tuned.[6,7] In a real system a number of factors like defects, confinement, and surface tension are important to understand pressure induced electronic and structural transformations. Therefore, the studies of low dimensional materials under pressure can provide valuable informations.[8] High pressure Raman scattering is a powerful technique for extracting information regarding vibrational properties and phase diagrams of both bulk and nanostructures.[6, 9-14] The use of diamond anvil cell and ruby fluorescence allows reaching very high pressure, which enables one to study many new phenomena related to the behavior of solids.[6, 15-18]

In InAs, the stable structural phase at bulk state is zinc blende (ZB) but the NWs often stabilize in wurtzite (WZ) structure. The phonon modes in the zone center of the two structures are different. In WZ structure the modes can be estimated by backfolding the phonon branches of ZB InAs along [111].[19] Using Raman spectroscopy the differences between the modes of WZ and ZB phase can be probed. The pressure induced phase transition in bulk InAs at 8.46 GPa was first reported by Minomura and Drickamer[20] from resistivity measurements. Using Raman spectroscopy Jayaraman et al., reported the occurrence of the metallization at 7.15 GPa.[21] Vohra et al, studied InAs under pressure up to 27 GPa and observed the structural sequence from ZB to rock salt and to β-Sn.[21] Cardona *et al*. studied volume dependent Raman frequency, transverse effective charge ($e^*_T$) and resonance behavior of Raman intensity in bulk InAs under hydrostatic pressure.[22] Recently, the studies of the influence of hydrostatic pressure on the semiconductor low dimensional systems have attracted extensive research interest due to their application in opto-electronics devices. Pressure dependent electronic band gap of the WZ InAs NWs has been studied by resonant Raman scattering technique.[18] Furthermore, in ZB GaAs NWs, the pressure dependent lattice constant, ionicity, and electron–phonon Fröhlich interactions have been investigated [6] while the pressure dependence of structural parameters of WZ InAs NWs has never been studied.

In the present work, we report high pressure response of predominant WZ InAs NWs, in which ZB structure is present as a stacking fault. We have obtained mode Grüneisen parameters (γ) for all observed phonon modes at ambient pressure and also the transverse effective charge ($e^*_T$) from the measured transverse optical (TO) and longitudinal optical (LO) frequency as a function of pressure. We found a clear indication of structural phase change in the NWs above 10.87 GPa. In addition, the pressure dependence of the Raman scattering intensity measured with a fixed laser line has been discussed in terms of resonant Raman scattering near the $E_1$ gap.

**RESULTS AND DISCUSSION**

Aligned InAs NWs were grown on a InAs (111)B substrate using chemical beam epitaxy (CBE) technique. The NWs were grown at 425 °C, with metal organic line pressure of 0.3 and 1.0 Torr for trimetyl indium (TMIn) and tertiarybutyl arsine (TBAs), respectively. The detailed fabrication technique is described elsewhere.[23] Shape, size and crystal structure of the NWs were studied using transmission electron microscopy (TEM, Model: JEOL 2010). The Raman measurements were performed in a diamond anvil cell (DAC) in back scattering geometry. A 4:1 mixture of methanol and ethanol was used as the pressure transmitting medium. Pressure was monitored in situ by the shift in the Ruby fluorescence peak. Argon ion laser was used as an excitation source and the spectra were collected with LabRam HR monochromator and CCD detector (Jobin Yvon).

Fig. 1 (a) and 1(b) show the low and high magnification TEM images of NWs. The stripes along the length of the NWs are the signature of the presence of ZB phase, which is present as staking faults in the WZ NWs. The average diameter of the NWs is 40 nm. Recent report on the same NWs shows that the percentage of WZ and ZB structures on the whole NW length are 95% and 5% respectively.[24] Fig. 1(c) displays the Raman spectrum of bulk InAs (ZB structure). The peak positions are determined by fitting the spectrum using Lorentzian line shapes. The positions of the zone center transverse optical (TO) and longitudinal optical (LO) phonon modes are observed at 217.65 cm$^{-1}$ and 235.33 cm$^{-1}$, respectively. The Raman spectrum of InAs NW at ambient pressure is shown in Fig. 1(d). The TO exhibits an asymmetric and broad peak compared to bulk InAs. We have carefully fitted the spectrum with four Lorentzian functions. The peak at about 217.9 cm$^{-1}$ is the TO mode (blue line) and the LO mode (green line) is appeared at 238.4 cm$^{-1}$. $E_2^H$ WZ mode (cyan line) and surface phonon (SO) mode (pink line) are observed at 214.2 and 231.23 cm$^{-1}$, respectively. Thus the Raman study enables us to confirm the WZ structure in the NWs.

Schematic of the high pressure Raman setup using diamond anvil cell is shown in the Fig. 2 (a). The Raman spectra of single InAs NW as a function of hydrostatic pressure are plotted in Fig. 2 (b). With the increase of pressure all phonon modes shift towards higher frequency side with gradually decreasing intensity and finally vanish above 10.87 GPa. The frequencies of the measured modes as a function of applied hydrostatic pressure and reduction in volume are shown in Figs. 3 (a) and 3 (b), respectively. Since the energy difference between $E_2^H$ and TO mode is small (~3.7 cm$^{-1}$), it is difficult to differentiate their individual pressure dependent behavior. Similar to Ref. 18 we have used single Lorentzian to fit the curve in the spectral range of the two modes ($E_2^H$ and TO) and obtained fitted parameters have been used for TO mode related calculations. The relative volume compression has been obtained from Murnaghan's equation:[25]

$$P(V) = \frac{K_0}{K_0'}[(\frac{V}{V_0})^{-K_0'} - 1]\ldots\ldots\ldots\ldots(1)$$

Where $K_0 = -V((\partial P)/(\partial V))$ is the isothermal bulk modulus, $K'_0$ its pressure derivative and $V/V_0$ is the volume reduction fraction with pressure, where $V$ and $V_0$ are the volumes corresponding to the pressure examined and at zero pressure, respectively. We have used the reported values of $K_0$ (59.60 GPa)[26] and $K'_0$ (4.62)[26] for WZ InAs to calculate volume reduction using Eq. 1.

In most of the ZB semiconductors, the pressure dependence of phonon frequencies is usually modeled by a quadratic equation. But in the present experiment the pressure dependent phonon modes are well described by linear equation

$$\omega_i = \omega_{i0} \pm a_i P \ldots\ldots\ldots\ldots(2)$$

where $\omega_{io}$ is the $i^{th}$ phonon frequency at zero pressure, $P$ is the pressure and $a_i = (\partial \omega_i / \partial P)_{P=0}$ is the linear pressure coefficient of $i^{th}$ mode. Such linear dependence on pressure of the phonon modes has also been observed for WZ InN semiconductor [27-29] and WZ ZnO.[30]

The dependence of the TO, SO, LO and 2LO phonon modes on pressure obtained from fitting of the experimental data are as follows.

$$\left. \begin{array}{c} \omega_{TO} = (215.8 \pm 0.9) + (3.75 \pm 0.13)P \\ \omega_{SO} = (234.3 \pm 1.6) + (2.91 \pm 0.23)P \end{array} \right\} \ldots\ldots\ldots\ldots (3)$$

$$\omega_{LO} = (239.5 \pm 0.8) + (3.23 \pm 0.12)P$$

$$\omega_{2LO} = (476.9 \pm 1.6) + (7.51 \pm 0.88)P$$

where frequencies ($\omega$) are in cm$^{-1}$ and pressure ($P$) is in GPa.

The variation of the phonon frequencies as a function of volume reduction (computed using Eq. 1) are as follows

$$\left.\begin{array}{l} \omega_{TO} = (547.8 \pm 7.3) - (334.1 \pm 7.8)\dfrac{V}{V_0} \\[4pt] \omega_{SO} = (494.1 \pm 10.5) - (261.6 \pm 11.3)\dfrac{V}{V_0} \\[4pt] \omega_{LO} = (526.1 \pm 4.7) - (288.5 \pm 5.0)\dfrac{V}{V_0} \\[4pt] \omega_{2LO} = (976.3 \pm 72.9) - (499.5 \pm 74.5)\dfrac{V}{V_0} \end{array}\right\} \quad \ldots\ldots\ldots (4)$$

The values of $\omega_{io}$ and $a_i$ are summarized in Table I. From the obtained data we computed the mode Grüneisen parameter, which is defined as the shift in phonon frequency as a function of pressure change and is given by [31,32]

$$\gamma_i = \frac{\partial \ln \omega_i}{\partial \ln V} = \frac{K_0}{\omega_{i0}} a_i \quad \ldots\ldots\ldots (5)$$

The obtained Grüneisen parameters given in Table I, are lower than the reported values for bulk InAs (for zone center TO and LO modes). The reduction of the estimated Grüneisen parameters may be due to the smaller variation of the phonon frequency with pressure in one dimensional systems. The defects within the NWs may also be the origin of the reduction in $\gamma_i$ values. It is also interesting to note that the values of $\gamma_i$ are close to 1, which might be signature of covalency in the WZ NW.[33]

Fig. 4 (a) shows the LO - TO splitting as a function of applied pressure. We found that the splitting decreases with increasing pressure and shows linear dependence ($\omega_{LO} - \omega_{TO} = (23.7 \pm 0.2) - (0.51 \pm 0.03)P$) on pressure. This result is consistent with previous observation in GaN, AlN, GaAs and GaP.[10, 12, 34]

Born's effective charge ($e^*_T$) is associated with the absorption induced by TO phonons. It is one of the dynamical effective charges, which is a measure of the ionicity of a compound. The relationship between $e^*_T$ and $\omega^2_{LO} - \omega^2_{TO}$ is given by[35]

$$(\omega^2_{LO} - \omega^2_{TO}) = \frac{(e^*_T)^2}{\varepsilon_0 \varepsilon_\infty \mu V} \dots\dots\dots\dots(6)$$

where, $\mu$ being the reduced mass of the anion-cation pair; V is the volume of the primitive cell, $\varepsilon_0$ is the vacuum permittivity, and $\varepsilon_\infty$ the infrared dielectric constant. The calculated $e^*_T$ (in units of elementary charge) is decreasing linearly with increasing pressure and is fitted with the following function as shown in Fig. 4 (b).

$$e^*_T = (2.46 \pm 0.01) - (27 \times 10^{-3} \pm 0.001)P \dots\dots\dots\dots(7)$$

Similar pressure dependence has been observed for $e^*_T$ in bulk InAs,[22] and in WZ ZnO.[30] The decrease of $e^*_T$ with pressure indicates that the system becomes more covalent with lattice compression.[6] Therefore, the resistivity of the NWs increases under high pressure.

The contour plot of the intensity of the observed Raman modes as a function of applied pressure has been demonstrated in Fig. 5 (a). The Raman spectrum of the WZ InAs NWs change drastically above 10.87 GPa. The spectrum does not show any clear signals of the TO, LO, SO and 2LO modes above 10.87 GPa (as shown in both Fig. 5 (a) and in the inset of the same) but some new modes (labeled by A, B, and C in Fig. 5 (a) and 5(b)) appear in the lower frequency of TO and LO modes. Such new modes have been observed in bulk WZ InN semiconductor as a characteristic of WZ to rock salt phase transition above 13.5 GPa.[28] In the present study, the disappearance of all the phonon modes (observed at ambient pressure) and the appearance of the new modes may be interpreted as structural phase transition from WZ to rock salt crystal structure. In bulk InAs the ZB to rock salt structure transformation has been observed at 7 GPa by Vohra et.al.[21] In this study we found that the phase transition pressure (11.6 ± 0.7 GPa) is higher than that for bulk InAs (ZB structure) with volume reduction factor ($V/V_0$)~0.86 (shown in the inset of Fig. 3 (b)). This may be due to the fact that our NWs have predominant WZ structure with only 5% ZB phase. Moreover, in NWs the surface area is higher than the bulk material and hence it stores much higher surface energy than bulk. Therefore, NWs need higher pressure than bulk materials to overcome the extra energy for realization of structural transformation. Further experimental studies are needed for the confirmation of phase transition.

Fig. 6 displays the variation of intensity and linewidth (FWHM) of TO, LO SO and 2LO modes with pressure. The intensity profile of the above modes shows resonant behavior around 1.64 GPa. At ambient pressure the $E_1$ gap of WZ InAs is 2.4 eV[18]. In bulk InAs (ZB structure) the energy gap increases with applying pressure.[22] Assuming that the pressure dependence of the $E_1$ gap of WZ InAs is the same as ZB InAs (0.074 eV/GPa)[36] the calculated $E_1$ gap at 1.64 GPa is 2.52 eV. This energy value is close to the incident photon energy (2.54 eV). Therefore, a resonant Raman scattering is expected around 1.64 GPa, which agrees well with the measured peak of the intensity profile. The asymmetry shape of the intensity profile with pressure may be due to strong absorption of the incident and scattered radiation in the low pressure side.[22] Also, the asymmetric behavior agrees well with the occurrence of second maxima at higher pressure as observed by Zardo et. al.[18] The widths of LO, SO and 2LO peak decrease up to pressure ~2 GPa and then increase. The increase in the widths may be due to effect of anharmonicity in the vibrational deformation potential where phonon decays into two lower energy phonons.[24] For TO mode (which is unresolved ZB like TO ($A_1 + E_1$) mode and $E_2^H$ WZ mode), the spectral line width continues to decrease up to ~8 GPa and then increases. Such behavior may be the effect of strain due to the presence of WZ and ZB interfaces along the length of the NWs.[24]

**CONCLUSIONS**

In summary, pressure dependent (up to ~58 GPa) phonon modes of InAs NWs with predomnant WZ structure (95%) are investigated by high pressure Raman measurements. The linear pressure coefficient and mode Grüneisen parameters of TO, LO, SO and 2LO modes have been calculated. The Grüneisen parameters are found to be close to one, which indicates that the system contains covalent bonds. The decrease of LO-TO splitting and Born's transverse effective charge with increasing pressure supports the reduction of ionicity and increase of covalency of the system. Our results are in good agreement with other III-V semiconductors.[6, 22, 27-29] The resonance behavior of the intensity of all the Raman modes up to ~ 2 GPa shows that the $E_1$ gap of WZ InAs NWs approaches the excitation energy (2.54 eV). In addition, an indication of structural phase transition of the predominant WZ NWs at 11.6 ±0.7 GPa has been observed. We propose that the transition is from WZ to rock salt structure but further experimental and theoretical investigations are needed to clarify this signature.

**ACKNOWLEDGMENTS**

The authors acknowledge useful discussions with Biswajit Karmakar, Saha Institute of Nuclear Physics, Kolkata, India. Authors also thank P.V. Satyam, IOP, Bhubaneshwar, India, for his help in the TEM work. The work was partly supported by MIUR under PRIN 2009 prot. 2009HS2F7N_003.

TABLE I: Results of linear regression of the pressure dependence of phonon modes ($\omega_{i0}$ and $a_i = \left(\frac{\partial \omega_i}{\partial P}\right)_{P=0}$) and calculated mode Grüneisen parameters ($\gamma_i$). Reported values of $a_i$ and $\gamma_i$ for bulk InAs are given in the last two columns.

| Phonon modes | Present work | | | Bulk InAs[a] | |
|---|---|---|---|---|---|
| | $\omega_{i0}$ (cm$^{-1}$) | $a_i$ (cm$^{-1}$GPa$^{-1}$) | $\gamma_i$ | $a_i$ (cm$^{-1}$GPa$^{-1}$) | $\gamma_i$ |
| TO | 215.8 | 3.75 | 1.04 | 4.39 | 1.21 |
| SO | 234.3 | 2.91 | 0.74 | | |
| LO | 239.5 | 3.23 | 0.80 | 4.57 | 1.06 |
| 2LO | 476.9 | 7.51 | 0.94 | | |

[a]From Ref. 22

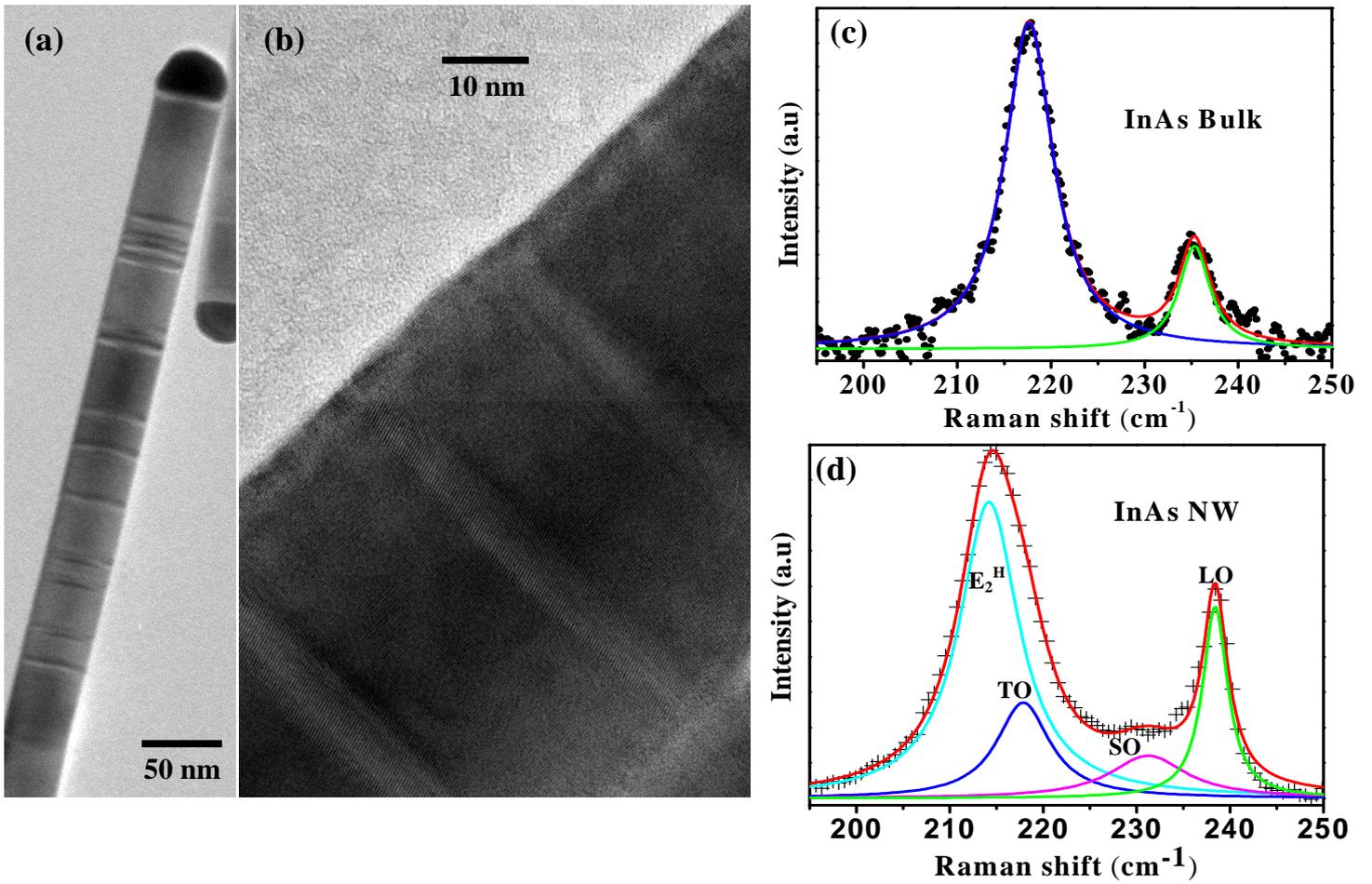

FIG. 1. TEM dark field images of InAs NWs (a) low magnification, (b) high magnification, (c) Raman spectrum of bulk InAs (symbols: ●); the red line is the result of a fit with two Lorentzian functions for TO (blue) and LO (green) phonons. (d) WZ InAs NWs Raman spectrum (symbols: +); the red line is the result of the fit with four Lorentzian components (see text).

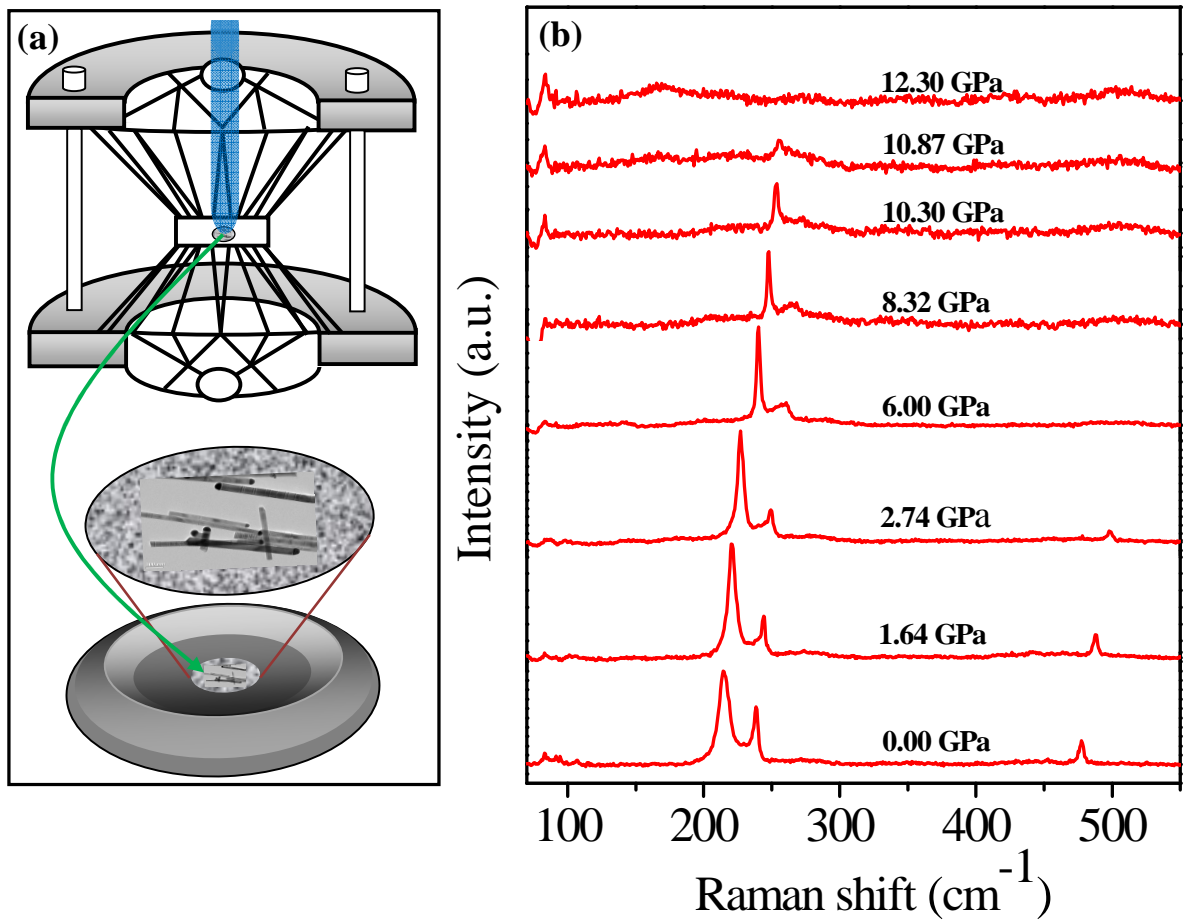

FIG. 2. (a) Schematic representation of the high pressure Raman measurement setup. The blue line represents both the incident beam and the backscattered signal. In the bottom, a magnified representation of the sample chamber with NWs inside DAC. (b) Pressure dependence of the Raman spectra from 0 to 12.30 GPa. For clarity, the spectra are shifted vertically.

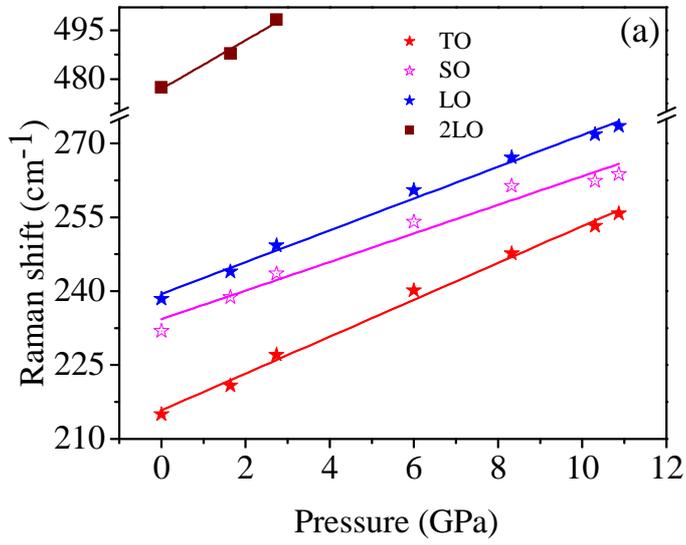 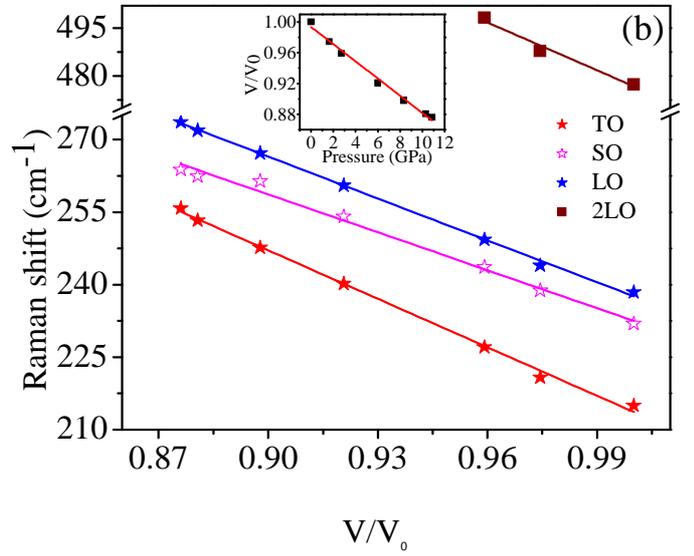

FIG. 3. Raman frequency as a function of (a) applied pressure, (b) volume reduction. The lines are the result of the linear fit according to Eq. 2. Inset of Fig. 3b shows the volume reduction factor as a function of applied pressure, obtained through Eq. 1.

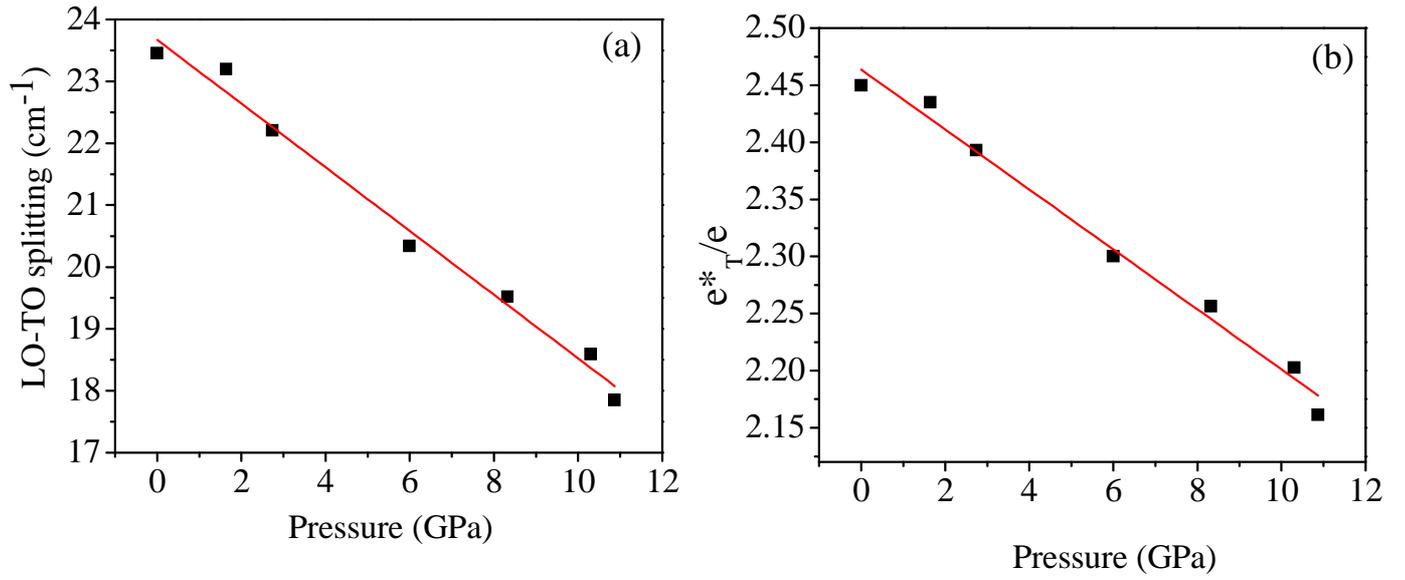

FIG. 4. (a) Splitting of the LO and TO modes as a function of pressure. (b) Born's effective dynamical charge as a function of pressure computed from the experimental data of Fig. 4a using Eq. 6. The solid lines in both figures are linear least-square fit to the data points.

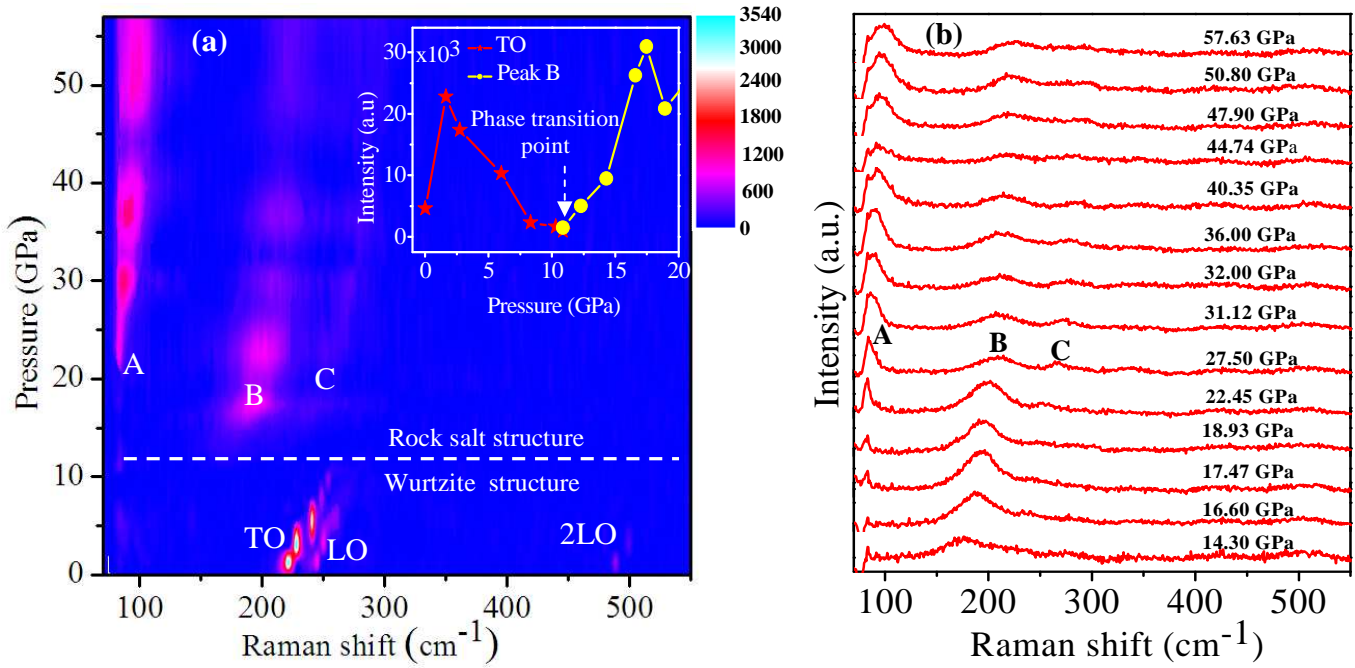

FIG. 5. (a) Contour plot of the intensity of the Raman modes as a function of pressure (0 to 57.63 GPa). The intensity variation of TO mode and Peak B as a function of applied pressure (up to 20 GPa) is shown in the inset of Fig. 5 (a). The white arrow indicates the phase transition point (11.6±0.7 GPa) at which TO peak vanishes and Peak B arises. (b) Pressure dependence of the Raman spectra from 14.30 to 57.63 GPa. For clarity, the spectra are shifted vertically

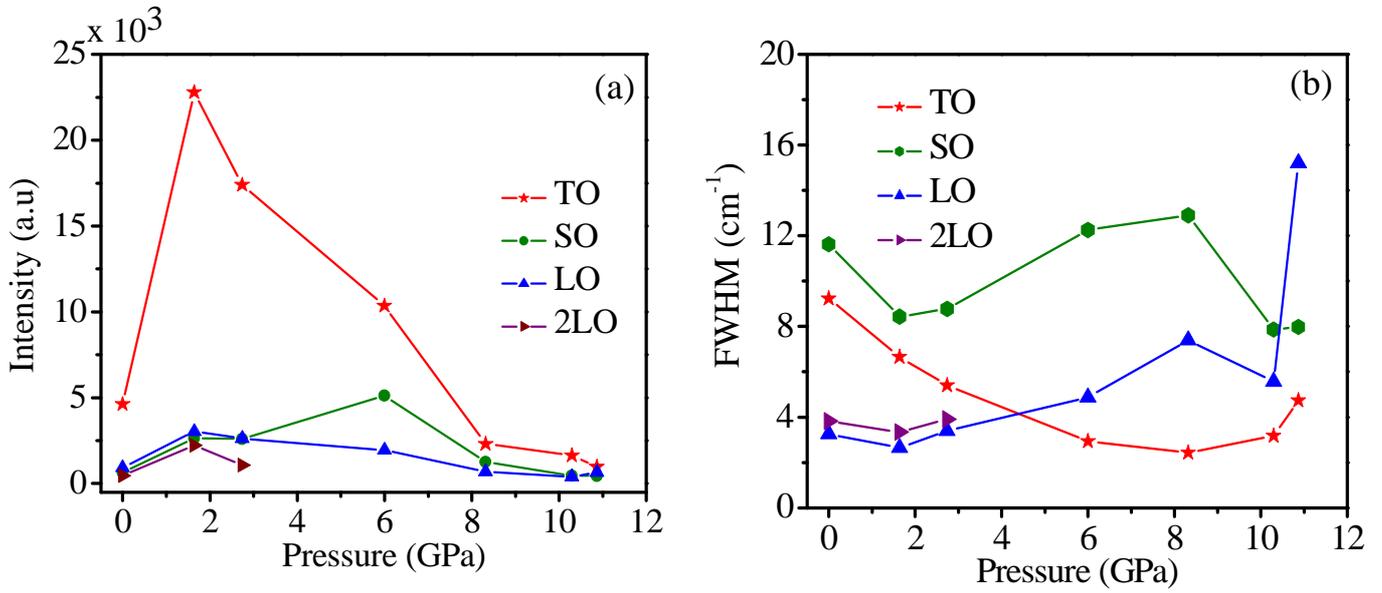

FIG. 6. (a) Intensity and (b) FWHM of the TO, SO, LO and 2LO modes as a function of the applied pressure.